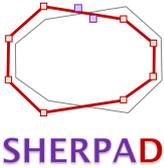

# SHERPAD: test of Slow high-efficiency extraction or Positrons from a Ring At DAFNE.

P. Valente[1], D. Annucci[1,2], O. R. Blanco Garcia[3,*], M. Garattini[3], P. Gianotti[3], S. Guiducci[3], A. Liedl[3], M. Raggi[1,2]

[1]INFN Sezione di Roma, P.le Aldo Moro 2, 00185 Rome, Italy
[2]Dipartimento di Fisica, "Sapienza" Università di Roma, P.le Aldo Moro 2, 00185 Rome, Italy.
[3]INFN Laboratori Nazionali di Frascati, Via Enrico Fermi, 00044 Frascati (Rome), Italy.
[*]now at Synchrotron SOLEIL, L'Ormes de Merisiers Saint-Aubin BP 48 91192 Gif-sur-Ivette Cedex, France.

## Abstract
The idea of using fixed-target annihilations of a high-energy positron beam on a target for producing a new, very feebly interacting particle has been exploited by the PADME experiment at LNF using the extracted beam from the LINAC in the BTF facility. Extracting the beam from a synchrotron would improve by several orders of magnitude the duty-cycle of the LINAC, thus greatly extending the physics reach of such experiments. The option of substituting or complementing the standard, slow 1/3 of integer resonant extraction technique by using coherent phenomena in a bent crystal, like the channeling has been studied by the SHERPA experiment. Both using the DAFNE main positron ring or the smaller damping ring (accumulator) as pulse extender of the LINAC have been considered.

We here propose to demonstrate the feasibility of this technique performing an experiment on the DAFNE positron main ring, installing in a suitable vacuum goniometer a thin Silicon crystal, bent in the (110) direction in order to provide a ~1 mrad kick, and then simulating the extraction process with a "virtual septum": instead of performing a real extraction, the crossing of the gap of a septum magnet is simulated by detecting positrons at its position (at phase advance close to 1/4) by means of a Silicon pixel detector inserted in the ring, in a secondary vacuum of a "Roman pot".

## 1. State of the art
The idea of using fixed-target annihilations of a high-energy positron beam on a target for producing a new, very feebly interacting particle [1] has been exploited by the PADME experiment at LNF [2] in the case of a thin target, and similar experiments using thick targets [3] or at future positron facilities [4] have been proposed. In addition, coherent processes in a regular crystalline lattice are much more efficient for positively charged particles, and in particular a positron beam interacting with a suitable bent crystal is an ideal tool for studying innovative radiation sources [5].
Unfortunately, positron beam facilities are quite rare [6] and recently the idea of extracting the beam from an existing synchrotron has been put forward [7,8] for the DAFNE complex at LNF, to improve by several orders of magnitude the duty-cycle of the LINAC, limited to $\sim 10^{-5}$ [9]. The option of substituting or complementing the standard, slow 1/3 of integer resonant extraction technique by using coherent phenomena in a bent crystal, like the channeling, was also put forward in Ref. [7], and has been studied with more details by the SHERPA CSN5 young researchers grant [10,11]. Both using the DAFNE main positron ring or the smaller damping ring (accumulator) as pulse extender of the LINAC have been considered.

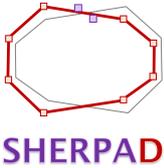

Steering of charged particle beams by means of coherent effects in crystals originally proposed by Tsyganov almost 50 years ago has been so widely studied and used in accelerators, especially for high-energy protons, to the point that it will be used for the beam halo collimation at the high-luminosity LHC (see Ref. [12] for a comprehensive review). However, channeling of low-momentum particles (0.5 GeV in the case of DAFNE) is affected by the Coulomb multiple scattering on the crystal itself, scaling as $p^{-1}$, thus demanding a very small thickness. On the other hand, even though the critical angle would be also larger at low momentum (scaling as $p^{-1/2}$) determining a larger angular acceptance for channeling, a thin crystal has to be bent to a relatively small radius in order to get a useful bending, of the order of a few mrad, sufficient for the beam extraction. Channeling of an 855 MeV electron beam was measured at MAMI in 15 µm and 30 µm thick Silicon crystals, bent by 315 µrad and 905 µrad respectively along (111) planes [13,14], while channeling of 480 MeV positrons in a 1 mm thick crystal, bent by ~10 mrad, has been barely observed at BTF [15].

The first objective of the SHERPA grant was to study a viable extraction scheme using a bent crystal in one of the DAFNE rings, considering the modifications to the machine and the requirements for the crystal itself i.e., its thickness, the position along the ring and with respect to the extraction septum, the required angular kick, etc. Once having assessed the crystal parameters, the second objective was to design and realize a working prototype of the crystal complete system, including the holder and bending system, as well as the goniometer for precise alignment with respect to the beam.

Realistic optics were studied both using the DAFNE main ring and the accumulator, leaving essentially unmodified the hardware layout of the machine. Particle in the MR+ losing a fraction $\Delta p/p = -0.7\%$ of their energy would reach a crystal positioned at $\Delta x$ =8 mm from the beam center providing a 1 mrad kick and would be extracted by a 2 mm thick septum at $\Delta x$ =20 mm placed at $\mu$ =0.192 in the same turn [16]. If the RF is left off and the momentum change is driven by synchrotron emission a spill of 300 µs will be produced. Similar results were obtained for the shorter damping ring, where the synchrotron losses are slightly different, yielding a spill of ~100 µs.

A 15 µm thick Silicon crystal was finally realized [17] and bent along the (110) direction using a dedicated motorized holder [18] by the INFN Ferrara group, and it is now ready to be tested and characterized on beam by the SHERPA collaboration, using a custom-realized vacuum chamber and a commercial goniometer. Also, the detector system, using a Timepix3 and a custom collimator, has been very recently tested with a 450 MeV beam from the BTF-2 line.

## 2. Objective

The objective of the SHERPAD project is to go a step further in the assessment of the feasibility of a crystal-assisted extraction of 0.5 GeV positrons from DAFNE, confronting all the issues related to the practical implementation of such a technique in a working synchrotron. The final aim is thus to be capable of performing an extraction test on the ring, and to characterize the extracted positron beam, in terms of extraction rate (intensity vs. time of the spill), and of the beam quality, in particular its emittance (size and divergence). Different schemes can be tested for producing the horizontal displacement necessary to reach the crystal, in order to get the highest and most uniform extraction rate: synchrotron losses, with or without a tune resonance, keeping the ring RF completely or partially off, etc. Moreover, the extraction can be optimized as a function of the machine parameters, like circulating current and optical functions, and of the crystal itself (orientation and position).

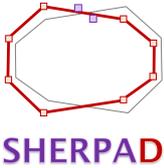

For reaching this objective a few basic elements are necessary, which will be the deliverables of the different work packages in the PBS:
- design and implementation of the machine layout and optics for the extraction;
- design and realization of the deflection system, including the crystal with its holder, the goniometer and the vacuum chamber;
- design and realization of a diagnostics system capable of detecting the "extracted" particle.

Concerning the last point, it is worth noticing that the realization of the real infrastructure like an extraction beamline from either of the two DAFNE rings (the main positron ring or the accumulator) would exceed the scope of this project, both from the point of view of the complexity and of the required resources.

A way of detecting and measuring the "extracted" beam using the existing magnetic and vacuum layout of the accelerator complex should then be devised, adapting and complementing the available diagnostics. The baseline is to prepare a test on the DAFNE positron ring, more suitable, with respect to the accumulator, in terms of available space, instrumentation, ancillary systems, access and safety, etc., since this would provide a validation of the calculations and simulations, which could then be extended to the accumulator option.

## 3. Methodology

The regions of the DAFNE positron main ring (MR+) for the location of the elements of the crystal extraction scheme can be identified based on the requirements of having the largest possible beta functions, which enhances the displacement $x_{ext}$ at the extraction point at a phase advance $\mu$ with respect to the kick $x'$: $x_{ext} = \sqrt{\beta_k \cdot \beta_{ext}} \cdot \sin(2\pi \cdot \mu) \cdot x'$. In addition, to have the smallest possible emittance of the beam, a small dispersion is required at the extraction.

In Fig. 1 two such regions are indicated: one close to the interaction point (IP) and one close to the diametrically opposite region where the otherwise completely separated electron and positron rings cross. Indeed, the two intertwined DAFNE "rings" have an octagonal shape, formed by two arcs with four 45° dipoles, one long "external" and one short "internal", connected by two straight sections. In the single interaction point configuration (implemented in the upgrade for the KLOE2 run), in the ring-crossing region (RCR) the two vacuum chambers are actually vertically separated.

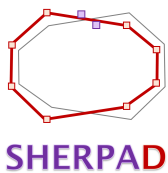

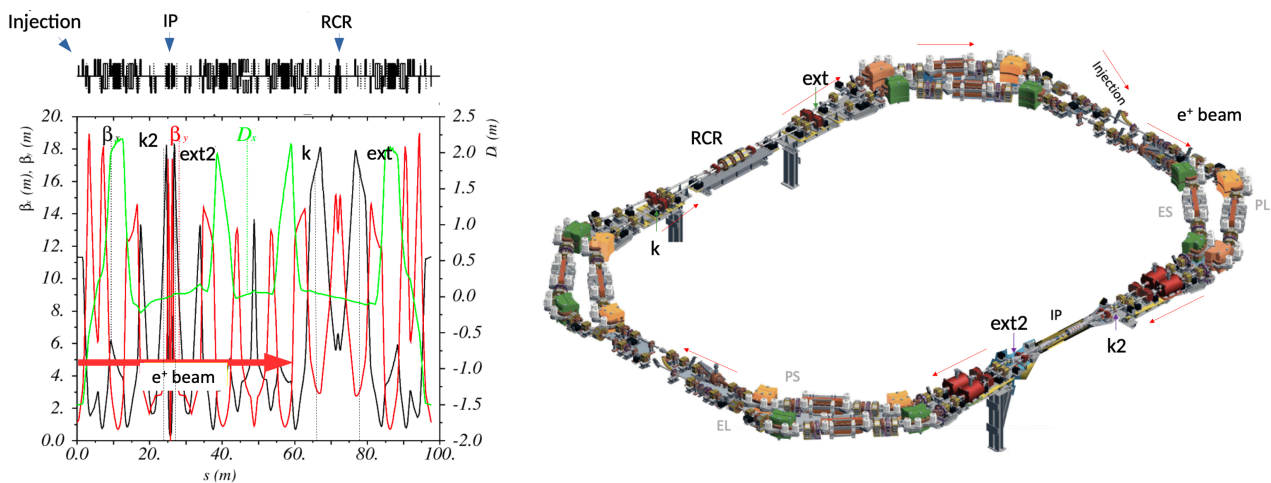

**Figure 1**: DAFNE positron main ring optical functions (SIDDHARTA configuration) and possible locations for crystal kick (k, k2) and extraction septum (ext, ext2), considering a kick along $x$.

In the actual extraction configuration, using the MR+ ring as pulse extender of the LINAC at the maximum repetition rate (49 Hz) would require modifying the injection scheme, producing the longest possible LINAC macro-bunch, in principle up to the length of the ring ~323 ns, and injecting it directly i.e. without passing through the accumulator. This is of course not necessary for running the test object of this project: the positron beam can be injected into MR+ from the accumulator, as in a standard collider injection sequence, in shorter bunches, <13.5 ns being the accumulator RF ~74 MHz. Then, different schemes can be studied, in order to produce the momentum offset necessary to drive the particles towards the crystal like synchrotron losses, but also going towards a 1/3 of integer tune resonance will be explored, using a "monochromatic" extraction scheme like the one described in [8]. The crystal will be thus reached after a number of turns determined by its horizontal offset with respect to the nominal orbit and will impart to positrons hitting it a kick of ~1 mrad, which should produce a displacement sufficient to cross the gap of a thin (few mm) magnetic septum after some oscillations either in the same turn ("local" extraction) or after a few revolutions.

Once more, for assessing the feasibility of the extraction, it is not necessary to have an actual septum and extraction beamline: a detector measuring the horizontal displacement of particles would act as a "virtual septum", giving information on how many particles would be extracted at that location. The two main requirements for such a device are:
- a spatial resolution below the mm, to measure the horizontal displacement as a function of the optics of the machine and the crystal orientation and position;
- a good sensitivity to low intensity beams, since only a fraction of the beam will start interacting with the crystal, moreover the efficiency of channeling has to be considered: in principle the capability of detecting single particles would be necessary.

The latter requirement is related to the number of circulating positrons, which will determine the number of particles expected to cross the virtual septum through an overall extraction efficiency. Contrary to the "real" extraction configuration, where the challenge will be producing long LINAC macro-bunches with momentum spread compatible with the direct injection, despite the SLED compression of the LINAC RF, for this project a simple scheme can be used using a single positron bunch injected from the DAFNE accumulator into the MR+ ring. The macro-bunch is injected off-center with respect to the beam-pipe and stored; then the RF can be turned off to let the beam reach the crystal, which is inserted at a given depth inside the ring. Finally, the distribution of particles at the septum location, before and after the interaction of the beam with the crystal, is studied.

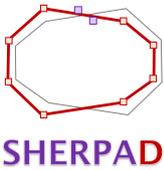

Assuming a momentum deviation of the order of $\Delta p/p = 1\%$, all the particles will reach the crystal in ~$10^3$ turns given the synchrotron loss of ~5 keV/turn (with wigglers off), corresponding to a "spill" of 0.3 ms. The efficiency of the channeling process for positively charged particles can be as high as 80%, provided that they impact the region with the correct curvature with an angle below the Lindhard critical angle, which is ~0.3 mrad for Si(110) at 0.5 GeV. The overall efficiency can however be much lower, considering several processes that can produce losses:
- "amorphous" interactions with the crystal of all particles hitting regions of the crystal with slightly different curvature;
- multiple interactions with the crystal of channeled positrons, if the virtual septum is not reached in a single turn;
- misalignment with respect to the nominal orbit, affecting the resulting deflection angle and acceptance;

In single bunch the design current of 40 mA has been exceed in both DAFNE rings, reaching more than 100 mA. Storing a current of 60 mA in a single bunch is thus realistic, corresponding to $10^9$ circulating positrons (1/RF=2.7 ns). However, at each turn only a small fraction of the beam will reach the crystal. DAFNE beam position monitors (BPM) have a resolution of ~20 $\mu$m above 3 mA and are sensitive down to 0.1 mA (with a resolution of order 100 $\mu$m) [19], so that they can be used for following the beam oscillations, but an additional imaging device is needed to measure the particles reaching the septum location.

This can be done inserting a finely segmented imaging device, like a Silicon pixel detector, inside the ring, either directly or in a secondary vacuum separated by a very thin window. For instance, TimePIX detectors with an array of 256×256 55 $\mu$m wide Silicon pixels have been routinely used in a secondary vacuum, in a so-called "Roman pot", for testing crystal collimation of the proton beam halo, both in the SPS and LHC rings [20].

## 4. Project organization
In the following, the project breakout system, organized in five work packages, is briefly described.

WP1, Machine layout: Design and implementation of the layout and optics of the ring.
The ring optics should be modified to optimize the optical functions at the crystal and at the virtual septum locations. The transverse distance of the crystal should be calculated, considering the oscillations amplitude, but also leaving enough aperture for injection. Finding a suitable configuration for the crystal and the (virtual) septum should be possible without significant modifications to the magnetic layout, as demonstrated by the preliminary SHERPA studies, in particular in the straight sections close the interaction point or the ring-crossing region (shown in Fig. 1). However, some adaptation to the pipes and vacuum system will be most probably needed for arranging the crystal assembly and the virtual septum: as an example, in Fig. 2 a detail of the RCR area is shown. The longitudinal and transverse space will be defined precisely by the WP4, but approximate dimensions will be sufficient for the definition of the layout.

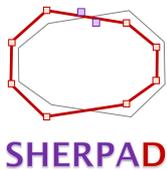

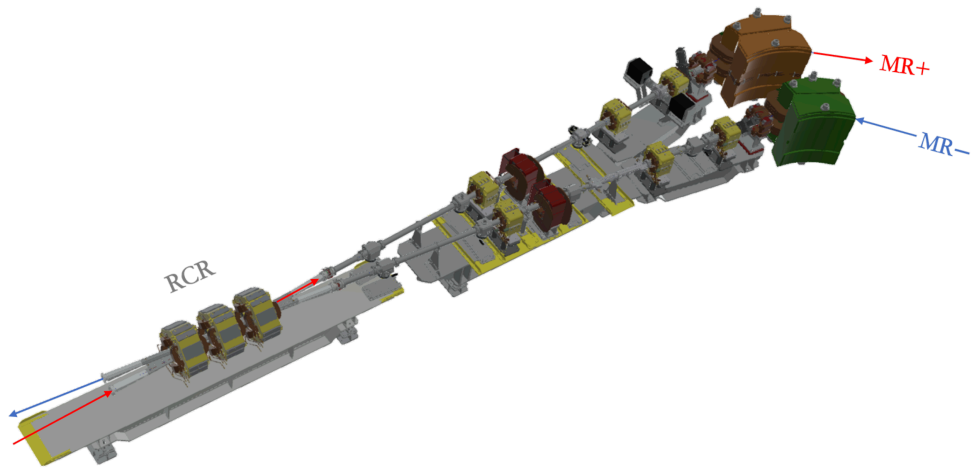

**Figure 2**: 3D drawing of the DAFNE MR+ main ring between the RCR and the long arc.

The optics of the ring has will be optimized for:
- placing the virtual septum at a phase advance of 1/4 with respect to the kick;
- have the highest possible beta functions in both locations;
- larger possible momentum acceptance;
- off-axis injection.

WP2, Crystal assembly: Design and realization of the crystal system.
The crystal cannot work without being integrated in an assembly, comprising its holder, which imparts the correct curvature, and a translational and rotation system (the so-called "goniometer") allowing to precisely align it to reach the channeling condition.
This project can profit of the experience gained in the context of the SHERPA grant, however some modifications have to be done in order to use a crystal+goniometer assembly inside the machine. The most relevant is that the holder should be modified in order to reach the beam periphery with the crystal edge without hitting first any part of the mechanics, which is clearly not the case with the design adopted for SHERPA.

WP3, Diagnostics: design and realization of diagnostic devices.
DAFNE BPM's are all along the ring, however additional devices or a different configuration of the existing ones could be needed, once the final layout will be finalized. In addition to BPM's, capable of following the orbit of the main part of the beam, a finely segmented imaging detector has to be inserted into the ring at the septum position, in order to detect particles that would be extracted and characterize them.
This detector could be places either directly inside the ring pipe or in a secondary vacuum separated by a thin window from the primary one. In the case of a Silicon pixel detector the latter solution has been routinely used, for instance TimePIX detectors with an array of 256×256 55 $\mu$m wide pixels was used in a secondary vacuum, a so-called Roman pot, both in the SPS and LHC rings [20].
The acquisition of the BPM's is integrated in the DAFNE control system, however retrieval and reduction of this data, as well as integration and synchronization with the information coming from the TimePIX will be necessary.

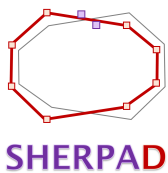

WP4, Vacuum.
The design of the vacuum devices, both the crystal assembly and the in-vacuum detector (the virtual septum) can have an impact on the layout of the ring. However, fixing the overall longitudinal and transverse maximum dimensions would allow the detailed design of the chambers to proceed minimizing the interference with the modifications to the layout and optics, even though some positions along the ring would not be usable due to the presence of supporting structures, cables and cooling pipes, ancillary systems, shielding, etc.
The crystal and holder produced by the SHERPA grant should be integrated in a complete system to be operated inside the machine vacuum. A custom vacuum chamber needs to be designed and realized, and a vacuum-compatible goniometer procured.
Concerning the virtual septum, both the detector case and the insertion vacuum chamber must be designed and built. Assuming to use a design similar concept to [21], a 0.38 mm thick steel window, ~2% of radiation length, should not spoil the spatial information on the very close TimePIX detector.

WP5, Machine tests and analysis. Performing the data taking, analysis and simulations validation. After installation of the crystal assembly and of the virtual septum, at least two data takings are foreseen, to leave the possibility of analyzing the data and in case correct possible issues in a second period of tests.

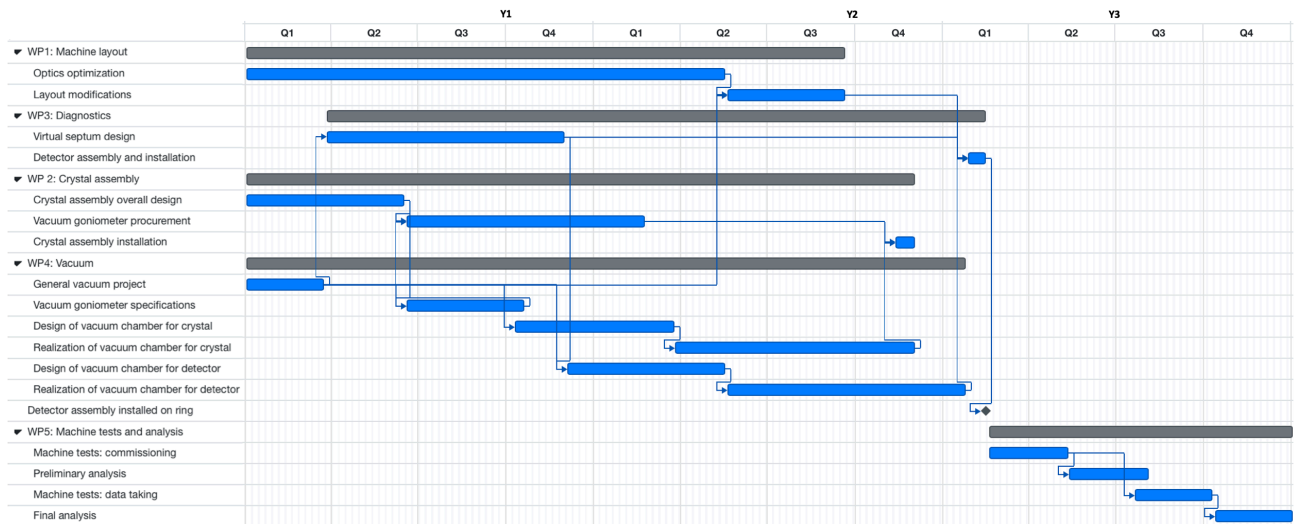

**Figure 5**: Tentative Gantt chart of the project.

## 5. Impact
Contrary to huge amount of information and applications on the channeling of protons and ions, there is much less experimental data on light particle channeling, like electrons and practically still no clean observation of positron channeling exists. The first successful extraction of positrons from a ring by means of coherent phenomena in crystals would be then a prime scientific result per se.

To this main, scientific impact, it should be added that a clear demonstration of an innovative extraction technique will open the possibility of realizing a facility. This is particularly interesting since, as already underlined, positrons beams are quite rare, and can have a variety of applications.

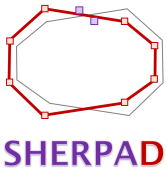

Indeed, the same crystal-assisted technique could be applied to similar machines, like the improvement of extraction efficiency for the DESY test beamline, which has been recently proposed [22].

Restricting to the DAFNE case, currently the LINAC can provide 0.5 GeV positrons limited to 49 Hz×300 ns which is a $1.5 \cdot 10^{-5}$ duty-cycle. Using the DAFNE main ring as pulse extender the duty cycle can be increased by three orders of magnitude. This would greatly extend the physics reach of the PADME experiment, currently running at the BTF-1 beamline.

A positron facility would have however a broader impact: since positively charged particles have a larger efficiency for coherent phenomena in crystals, thanks to their small mass, positrons are of particular interest for the development of innovative methods of radiation production using channeling.

## References

[1] B. Wojtsekhowski, D. Nikolenko and I. Rachek, arXiv:1207.5089.
[2] M. Raggi, V. Kozhuharov and P. Valente, EPJ Web Conf. 96 (2015) 01025.
[3] L. Marsicano et al., Phys. Rev. D98 (2018) 1, 015031.
[4] M. Battaglieri et al., Eur. Phys. J. A57 (2021) 8, 253.
[5] G. B. Susko, A. V. Korol and A. V. Solov'yov, arXiv:2110.13030 and references therein.
[6] P. Valente, EPJ Web Conf. 142 (2017) 01028.
[7] P. Valente, arXiv:1711.06877
[8] S. Guiducci et al., J. Phys. Conf. Ser. 1067 (2018) 6, 062006.
[9] P. Valente et al., J. Phys. Conf. Ser. 874 (2017) 1, 012017.
[10] M. Garattini et al., PoS EPS-HEP2021 (2022) 878.
[11] M. Garattini et al., PoS PANIC2021 (2022) 080.
[12] W. Scandale et al., Int. J. Mod. Phys. A37 (2022) 13, 2230004.
[13] A. I. Sytov et al., Eur. Phys. J. C77 (2017) 901.
[14] A. Mazzolari et al., Phys. Rev. Lett. 112 (2014) 135503.
[15] S. Bellucci et al., Nucl. Instrum. Meth. B252 (2006) 3.
[16] M. Garattini et al., Phys. Rev. Accel. Beams 25 (2022) 3, 033501.
[17] A. Mazzolari, private communication.
[18] D. De Salvador et al., 2018 JINST 13 C04006.
[19] A. Stella et al., AIP Conference Proceedings 451 (1998) 378.
[20] W. Scandale et al., JINST 6 (2011) T10002.
[21] L. Pellegrino and G. Sensolini, Internal note LNF, ACCDIV-01-2020.
[22] A. I. Sytov et al., Eur. Phys. J. C82 (2022) 3, 187.